\documentclass[twocolumn,secnumarabic,amssymb, nobibnotes, aps, prd]{revtex4}
\usepackage{graphicx}

\usepackage{epstopdf}

\begin{document}

\title{Trap Distribution and the Impact of Oxygen-induced Traps on the Charge Transport in Poly(3-Hexylthiophene)}

\author{Julia Schafferhans $^1$}\email[Electronic mail: ]{julia.schafferhans@physik.uni-wuerzburg.de}
\author{Andreas Baumann $^1$}
\author{Carsten Deibel $^1$}
\author{Vladimir Dyakonov $^{1,2}$}
\email[Electronic mail: ]{dyakonov@physik.uni-wuerzburg.de}
\affiliation{$^1$ Experimental Physics VI, Faculty of Physics and Astronomy, Julius-Maximilians-University of W\"urzburg, Am Hubland, 97074 W\"urzburg, Germany}
\affiliation{$^2$ ZAE Bayern, Bavarian Center of Applied Energy Research, Am Hubland, 97074 W\"urzburg, Germany}

\begin{abstract}
The trap distribution in the conjugated polymer poly(3-hexylthiophene) was investigated by fractional thermally stimulated current measurements. Two defect states with activation energies of about 50~meV and 105~meV and Gaussian energy distributions were revealed. The first is assigned to the tail of the intrinsic density of states, whereas the concentration of second trap is directly related to oxygen exposure. The impact of the oxygen induced traps on the charge transport was examined by performing photo-induced charge carrier extraction by linearly increasing voltage measurements, that exhibited a strong decrease in the mobility with air exposure time.
\end{abstract}

\maketitle

\begin{figure}	[htb]
	\centering
	\includegraphics{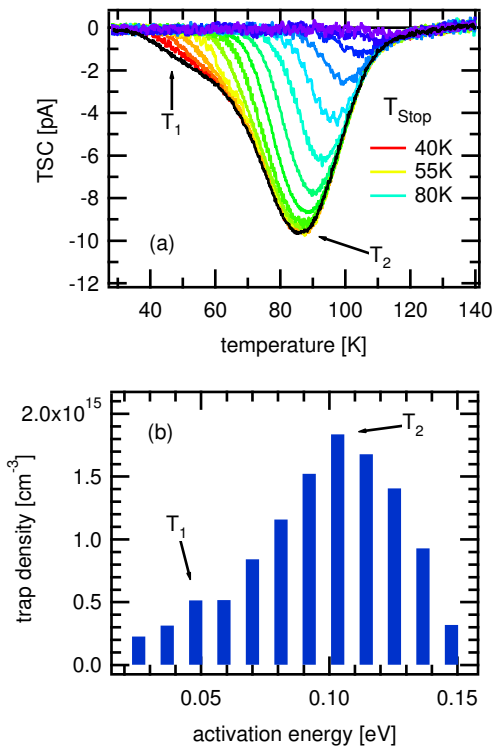}		
	\caption{(a) Main runs of the different $T_{Start}$- $T_{Stop}$ cycles, as well as the conventional TSC spectrum (black curve), revealing two trap states $T_1$ and $T_2$ and (b) the resulting DOOS distribution.}
	\label{fig:Fig1}
\end{figure}

Conjugated polymers are grown in interest for application in organic electronic devices such as organic light emitting diodes (OLEDs), field effect transistors and solar cells, due to their low cost processability from solution phase. The presence of traps can be critical to the performance of these devices since they reduce the charge carrier mobility, affect the driving voltage, disturb the internal field distribution and reduce the operation stability as well as the electroluminescence efficiency\cite{kadashchuk2003}. Concerning the lifetime of the devices, the influence of oxygen-related defect states on the charge transport might be decisive with respect to the long-term stability. In this work we investigated the trap distribution in poly(3-hexylthiophene) (P3HT) by a fractional thermally stimulated current technique. Furthermore the influence of oxygen on the trap states and their impact on the charge carrier mobility were examined.
\\
Diodes with the sandwich structure indium tin oxide (ITO)/poly(3,4-ethylenedioxythiophene) poly(styrene-sulfonate) (PEDOT:PSS)/ P3HT/Al were fabricated in nitrogen atmosphere. P3HT was spincoated on PEDOT:PSS covered ITO glass substrate, from 2~wt\% chlorobenzene solution at 800~rpm for 60~s, resulting in a film thickness of 220~nm. The 100~nm thick Al electrodes were thermally deposited at a deposition rate of 0.7~nm/s. Thermally stimulated current (TSC) and  photocharge extraction by linearly increasing voltage (photo-CELIV) measurements  \cite{juska2000, juska2003} were performed in a closed cycle cryostate (Janis CCS 550) with He atmosphere as contact gas. During the transfer to the cryostate, the samples were exposed to air for about five minutes. To optain the TSC spectra the samples were cooled down to 28~K. Trap filling was achieved by illumination of the samples with a 150~W halogen lamp for five minutes, since longer excitation times did not  show an increase in the TSC signal. After a dwell time of also five minutes the temperature was increased with a constant heating rate of 7~K/min up to 300~K.
During the measurements no external electric field was applied to avoid an overlap of the thermally stimulated and injection current, implying that the detrapped charge carriers were extracted from the sample only due to the built-in voltage, given by the work function difference of the electrodes. The TSC signals were detected with a source measurement unit (Keithley 237 SMU). Photo-CELIV measurements were performed at 300K. A $N_2$ laser with dye unit ($\lambda = 500~nm$) was used to generate charge carriers, which were then extracted by a linearly increasing reverse bias pulse.  The delay time between the laser flash and the triangular bias pulse was $30~\mu s$ and $80 ~\mu s$. The peak voltage in reverse direction was $V_a = 2~V$ applied on the Al electrode, with varied offset bias $V_{off}$ in forward direction. Pulse widths of $t_p = 100~ \mu s$  and $t_p = 500~ \mu s$ were chosen.
\\
A TSC peak between 40~K and about 130~K was detected in good agreement with the TSC measurements by Nikitenko et al. \cite{nikitenko2003}, but unlike their results no high temperature peak was observed. Due to the quite broad TSC peak, a distribution of trap states instead of a discrete trap level can be anticipated. To gain information on the trap distribution we applied a fractional TSC measurement, the so-called $T_{Start}-T_{Stop}$  technique \cite{schmechel2004,steiger2002}. It was performed as follows: trap filling was done at 28~K  ($T_{Start}$) as in the standard TSC technique, linear heating of the sample up to a temperature $T_{Stop}$ (\textit{prerelease}), subsequent cooling back to $T_{Start}$, and finally performing the whole TSC scan without  any new trapfilling (\textit{main run}). The whole procedure was repeated for increasing $T_{Stop}$, varied between 40~K and 105~K in steps of 5~K. The main runs for the different $T_{Stop}$ are shown in Fig.~\ref{fig:Fig1}(a). The sample exhibited a good stability without any aging effects during fractional TSC measurements, as is revealed by the fact that the main runs for the different $T_{Stop}$ finally merge perfectly into the conventional TSC spectrum measured before the fractional TSC (Fig.~\ref{fig:Fig1}(a), black curve). 
Additionally, a second conventional TSC measurement was performed after the whole $T_{Start}-T_{Stop}$ procedure, almost identical to the first run. The conventional TSC signal in the range between 26~K and 140~K exhibits a main TSC peak at about 85~K ($T_2$) as well as a shoulder at about 50~K ($T_1$), indicating two different trap distributions. In the fractional TSC spectrum the shoulder decreases for increasing $T_{Stop}$ due to the emptying of the shallower traps during the prerelease; it disappears for  $T_{Stop}$ larger than 60~K. The activation energies $E_t$ of the traps can be estimated for each $T_{Stop}$ interval from the initial rise regime of the respective main run, 
\begin{equation}
	\centering
		I_{TSC} \propto \exp \left(\frac{-E_{t}}{kT} \right)~~,
	\label{eq:eact}
\end{equation}
 where $I_{TSC}$ is the thermally stimulated current in the inital rise regime, $k$ is the Boltzmann constant and $T$ is the temperature. Hence, the activation energies for the $T_{Stop}$  intervals from 40~K to 95~K  were determined from the Arrhenius plot in the initial rise regime, whereas the signal-to-noise ratio was too low for the cycles with $T_{Stop}$ above 95~K. Thereby, a linear increase of the activation energy with the temperature was revealed, which has already been reported for fractional thermally stimulated luminescence measurements on different polymers \cite{kadashchuk1998,kadashchuk1999,kadashchuk2002,kadashchuk2003,kadashchuk2005}. This continuous increase of the activation energies with temperature without any plateaus indicates a continuous trap distribution instead of discrete trap levels \cite{malm2002}, as already expected on the basis of the broad TSC spectrum.  The lower limit of the trap density $n_{t}$ can be obtained by integrating the TSC spectrum over time, according to the inequality \cite{kadashchuk2005}
 \begin{equation}
	\centering
		\int_{peak} I_{TSC} dt \le e  n_{t} V  ~~,
 	\label{eq:numtraps}
\end{equation}
where $e$ is the elementary charge and $V$ is the volume of the sample. In this manner a lower estimation of the total ($T_1 + T_2$) trap density can be given, yielding $n_{t} = (1.0 -1.2)\times10^{16}~cm^{-3}$ for the investigated samples. Applying equation (\ref{eq:numtraps}) to the fractional TSC measurements, the trap densities for each $T_{Stop}$ interval can be estimated. These can be related to the activation energies extracted from the initial rise regimes, as shown in Fig.~\ref{fig:Fig1}(b), yielding a reconstruction of the density of occupied states (DOOS) \cite{schmechel2004}.
\begin{figure}[htb]		
	\centering
	\includegraphics{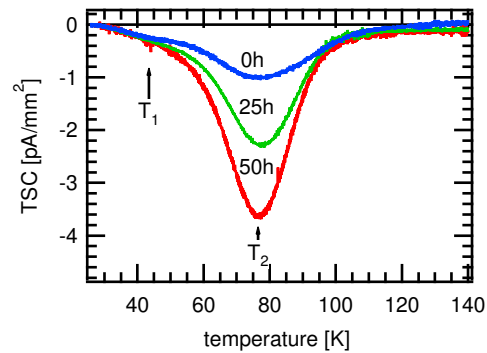}
	\caption{TSC spectra for different exposure times of the samples to air.}
	\label{fig:Fig2}
\end{figure}
The diagram displays a trap distribution with activation energies for the traps between 0.02~eV and 0.15~eV and two density maxima at about 50~meV and 105~meV, indicating two different traps ($T_1$, $T_2$) with approximately Gaussian energy distributions, in accordance with the observed shape of the TSC signal. The trap distribution $T_1$ can be attributed to the tail states of the intrinsic density of states (DOS) due to the low activation energy of a few 10~meV. $T_2$, however, shows a strong dependence on the exposure to air (Fig.~\ref{fig:Fig2}), 
\begin{table}[hb]
	\caption{Lower limit of the total trap densities obtained from the TSC measurements for different exposure times of the samples to air and dry oxygen.}
	\begin{center}
	\begin{ruledtabular}
	\begin{tabular}{lp{2in}}
	Sample & Total trap density [cm$^{-3}$] \\
	\hline
	Pristine & $1.0 \times 10^{16}$ \\
	25 h exposed to air &	 $1.7 \times 10^{16}$ \\
	50 h exposed to air & 	$2.6 \times 10^{16}$ \\
	\hline
	96 h exposed to O$_2$ & 	$2.8 \times 10^{16}$ \\
	\end{tabular}
	\end{ruledtabular}
	\end{center}
	\label{tab:1}
\end{table}
as was revealed by systematic exposure of the samples to air for several hours. 
\begin{figure}[htb]				
	\centering
	\includegraphics{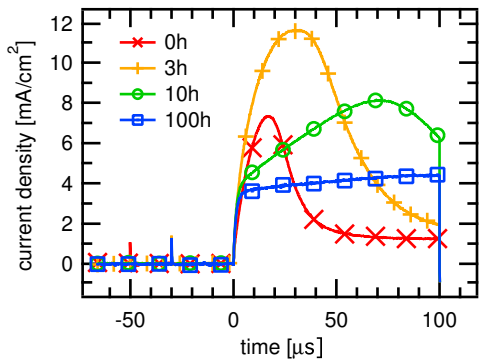}
	\caption{An exemplary photo-CELIV measurement ($V_{off} = -1.4~V$, $t_p = 100~\mu s$), showing the shift of the CELIV peak maximum to longer times with increasing exposure time of the sample to air, resulting in a decrease of the mobility.  }
	\label{fig:Fig3}
\end{figure}
With increasing exposure time, the TSC peak $T_2$ strongly increases, signifying a strong raise of the trap density of the deeper trap. In contrast, the shoulder $T_1$ in the TSC spectrum of the pristine sample, identified as the tail states of the intrinsic DOS, is unaffected by the exposure to air. For increasing exposure time, the shoulder seems to vanish as an effect of the overlap with the strong increasing trap $T_2$. To distinguish whether the increasing density of the deep trap is due to oxygen or moisture, a sample was exposed to pure oxygen, resulting in the same effect as for the exposure to air. Therefore, the strong increase of $T_2$ can be attributed to oxygen instead of moisture. It has to be mentioned that the exposure to air and oxygen did not result in any additional TSC peaks at higher temperatures than 130~K. The resulting total trap densities for the several exposure times are summarized in Table~\ref{tab:1}. After an exposure of 50~h to air, the total trap density yields $n_{t} = 2.6 \times 10^{16}~cm^{-3}$, which is more than two times higher as in the pristine samples. For comparison, the charge carrier density in OLEDs is also just in the range of  $ 10^{15} - 10^{16}~cm^{-3}$ \cite{tanase2003}. Therefore, an impact of the oxygen induced defect states on the performance of OLEDs can be expected. If the deeper traps are solely caused by oxygen, or only intensified due to the oxygen exposure, can not be answered yet, since also the pristine samples were exposed to air for a few minutes during the transfer to the cryostate, as mentioned before.\\
To investigate the impact of the oxygen induced traps on the charge carrier mobility, photo-CELIV measurements were performed. Therefore the samples were also exposed to air for several hours as in the TSC measurements. Since CELIV is a charge extraction technique, potential changes of the injection barrier due to air exposure will not strongly influence the mobility measurement. In Fig.~\ref{fig:Fig3}  photo-CELIV measurements on a P3HT sample for different exposure times are shown. The mobility was calculated from the position of the CELIV peak maximum \cite{juska2000}. From Fig.~\ref{fig:Fig3} it can be seen that with increasing exposure time, the CELIV peak shifts to longer times resulting in a lower mobility. The measured mobilities for the different exposure times of the samples to air are summarized in Table~\ref{tab:2}. After 100~h in air, the mobility decreases about two orders of magnitude as compared to the pristine samples. This effect can be correlated to the strong increase of the deep traps density. 
\begin{table}[htb]
	\caption{Charge carrier mobility obtained from the photo-CELIV measurements for different exposure times of the samples to air and different offset voltages applied.}
	\begin{center}
	\begin{ruledtabular}
	\begin{tabular}{lll}
	Sample &  Mobility [cm$^{2}/Vs$]  \\
	 & $V_{off}= -1.4~V\footnote{Underestimated mobility, due to high RC constant}$ & $V_{off}= -0.4~V$ \\
	 \hline
	Pristine & $2.5 \times 10^{-6}$  & $1.2 \times 10^{-4}$ \\
	3 h exposed to air & $5.2 \times 10^{-7}$  & $2.6 \times 10^{-5}$ \\
	10 h exposed to air & $1.2 \times 10^{-7}$ & $4.9 \times 10^{-6}$ \\
	100 h exposed to air & $5.4 \times 10^{-8}$ & --- \footnote{No estimation possible, due to low signal}\\
	\end{tabular}
	\end{ruledtabular}
	\end{center}
	\label{tab:2}
\end{table}
\\ In summary, we investigated the trap distribution in P3HT by a fractional stimulated current technique. Two different trap states were revealed, with activation energies of 50~meV and 105~meV and Gaussian energy distributions. The first can be assigned to the tail states of the intrinsic DOS due to their low activation energy. The second, however, is strongly influenced by exposure to oxygen. The lower limit of the total trap density shielded $n_{t} = 1.0 \times 10^{16}~cm^{-3}$ for the pristine sample, which rises up to $2.6 \times10^{16}~cm^{-3}$ after an exposure for 50~h to air. We demonstrated that the charge carrier mobility in our devices decreases about two orders of magnitude, already after an exposure to air for 100~h due to the strong rise of the deep traps density.

\end{document}